\documentclass{aa}
\usepackage{graphicx}
\newcommand{\beq}{\begin{equation}}
\newcommand{\beqa}{\begin{eqnarray}}
\newcommand{\eeq}{\end{equation}}
\newcommand{\eeqa}{\end{eqnarray}}
\newcommand{\siml}{\la}
\newcommand{\simg}{\ga}
\newcommand{\meszaros}{M${\acute {\rm e}}$sz${\acute {\rm a}}$ros}

\begin{document}
\authorrunning{K. Ioka, K. Toma, R. Yamazaki, \and T. Nakamura}
\titlerunning{Efficiency crisis of GRBs: prior activity or microphysics?}

\title{Efficiency crisis of {\it Swift} gamma-ray bursts with shallow
X-ray afterglows: prior activity or time-dependent microphysics?}

%\subtitle{}

\author{K. Ioka \inst{1}, K. Toma \inst{1}, 
R. Yamazaki \inst{2}, \and T. Nakamura \inst{1}}

\offprints{K. Ioka \email{ioka@tap.scphys.kyoto-u.ac.jp}}

\institute{Department of Physics, Kyoto University, Kyoto 606-8502,
Japan
\and
Department of Physics, Hiroshima University, Higashi-Hiroshima 739-8526,
Japan
}

\date{Received; accepted}

\abstract
{Most X-ray afterglows of gamma-ray bursts (GRBs) observed by
the {\it Swift}{} satellite have a shallow decay phase $\propto
t^{-1/2}$ in the first few hours.}
{This is not predicted by the standard afterglow model
and needs an explanation.}
{We discuss that the shallow
decay requires an unreasonably high gamma-ray efficiency, $\simg
75$--$90\%$, within current models, which is difficult to produce by
internal shocks.
Such a crisis may be avoided if a weak relativistic
explosion occurs $\sim 10^3$--$10^6$ s prior to the main burst 
or if the microphysical parameter of the electron energy
increases during the shallow
decay, $\epsilon_e \propto t^{1/2}$.
The former explanation predicts a very
long precursor, while both prefer dim optical flashes from
the reverse shock, as was recently reported.
We also calculate the multi-wavelength afterglows and compare them
with observations.}
{No optical break at the end of the shallow X-ray decay
indicates a preference for the time-dependent microphysics model
with additionally decaying
magnetic fields, $\epsilon_B \propto t^{-0.6}$.}
{}


\maketitle

%
%________________________________________________________________

\section{Introduction}

Recently the {\it Swift} satellite has allowed us to
observe early afterglows of gamma-ray bursts (GRBs)
in the first few hours after the burst
(e.g., Tagliaferri et al. \cite{taglia05};
Burrows et al. \cite{burrows05};
Chincarini et al. \cite{chinca05};
Nousek et al. \cite{nousek06};
Cusumano et al. \cite{cusuma06};
Hill et al. \cite{hill06};
Vaughan et al. \cite{vaughan06};
Barthelmy et al. \cite{barthelmy05}).
This time window remains largely unexplored,
and multi-wavelength studies of early afterglows 
would reveal many questions concerning GRBs,
such as the emission mechanism,
nature of the central engine,
and burst environment
(Zhang et al. \cite{zhang06};
Nousek et al. \cite{nousek06};
Yamazaki et al. \cite{yama06};
Toma et al. \cite{toma06};
Kobayashi et al. \cite{koba05};
Panaitescu et al. \cite{pana06};
Eichler \& Granot \cite{eichler06};
Granot \& Kumar \cite{granot06a};
Lazzati \& Begelman \cite{lb05}).

Early X-ray afterglows observed by the {\it Swift} X-Ray Telescope (XRT)
have three kinds of canonical features that are not predicted by
the standard model from the pre-{\it Swift} era 
(Nousek et al. \cite{nousek06}; Zhang et al. \cite{zhang06}).
X-ray light curves show (i) a very steep initial decay 
($\propto t^{-\alpha_1}$ with $3 \siml \alpha_1 \siml 5$)
followed by (ii) a very shallow decay 
($\propto t^{-\alpha_2}$ with $0.2 \siml \alpha_2 \siml 0.8$)
that connects to the conventional late afterglow,
while about half of the afterglows have (iii) strong, rapid X-ray flares
minutes to days after the burst
(Piro et al. \cite{piro05};
Burrows et al. \cite{burrows05};
Ioka, Kobayashi \& Zhang \cite{ikz05};
Fan \& Wei \cite{fan05}).

The steep decay component is most likely the tail emission of the prompt
GRBs and/or of the X-ray flares 
(Nousek et al. \cite{nousek06}; Zhang et al. \cite{zhang06};
Yamazaki et al. \cite{yama06}).
Even if the emitting surface stops shining, 
we continue to see photons coming from the region
at large angles relative to our line-of-sight
because the emitting surface has a curvature.
Most photons from the large angles are not emitted in our
direction because of relativistic beaming, so that
the flux decays steeply.
Since the emission region moves outward on the surface,
the tail emission features, e.g., the decay index and smoothness,
would diagnose the unknown GRB jet structure 
(Yamazaki et al. \cite{yama06}).

The X-ray flares are considered to be produced by the 
long activity of the central engine up to the time of the flares
(Burrows et al. \cite{burrows05}; Ioka et al. \cite{ikz05}).
This is mainly because an afterglow cannot be
variable with a large amplitude and a short timescale
by itself, i.e., such as by the ambient density fluctuations
and the inhomogeneous emitting surface,
as concluded by the kinematic arguments (Ioka et al. \cite{ikz05}).
However, the actual origin of the long activity is still under
investigation
(King et al. \cite{king05};
Perna, Armitage, \& Zhang \cite{perna06}).

The most enigmatic feature in early X-ray afterglows is 
the shallow decay of the light curve.
So far two kinds of models are proposed for the shallow X-ray afterglows.
One class of the models is the energy injection model
(Nousek et al. \cite{nousek06}; Zhang et al. \cite{zhang06};
Granot \& Kumar \cite{granot06a}),
in which continuous energy is injected into the afterglow
so that the flux decay becomes slower than the usual $\propto t^{-1}$.
The injection may be caused by (a1) the long-lived central engine 
(Dai \& Lu \cite{dai98}; Rees \& {\meszaros} \cite{rees00};
Zhang \& {\meszaros} \cite{zhang02})
or (a2) the short-lived central engine ejecting 
shells with some ranges of Lorentz factors
(Rees \& {\meszaros} \cite{rees98}; Kumar \& Piran \cite{kumar00};
Sari \& {\meszaros} \cite{sari00}; Zhang \& {\meszaros} \cite{zhang02}).
The other class is (b) the inhomogeneous jet model 
(Toma et al. \cite{toma06}; Eichler \& Granot \cite{eichler06}).
In this model, early afterglows are not bright because
the jet surface on the line-of-sight is dim
and the surrounding off-axis region with ordinary brightness
is observed later.

However, in all models,
the shallow X-ray afterglows pose a serious problem, demanding 
an unreasonably high gamma-ray efficiency of the prompt GRBs
(defined by $\epsilon_{\gamma} \equiv
E_{\gamma}/(E_{\gamma}+E_{k})$,
where $E_{\gamma}$ is the radiated prompt energy
and $E_{k}$ is the kinetic energy of the afterglow 
remaining after the burst),
as explained in Sect.~\ref{sec:crisis}.
Even before the {\it Swift} era,
one considers that
the gamma-ray efficiency of the prompt GRBs is relatively high,
i.e., $\epsilon_{\gamma} \sim 50\%$ or more 
(Lloyd-Ronning \& Zhang \cite{lloyd04}),
and develops internal shock models that 
can manage to produce such a high efficiency 
(Beloborodov \cite{belo00}; Kobayashi \& Sari \cite{koba01}).
Since the required efficiency is further increased,
we have a strong theoretical motivation to suspect
the current models.

In this Letter we suggest two more kinds of possible models for 
shallow X-ray afterglows without invoking an
unreasonably high gamma-ray efficiency.
One is the prior activity model in Sect.~\ref{sec:prior},
while the other is the time-dependent microphysics model in
Sect.~\ref{sec:micro}.
The energetic afterglow model in Sect.~\ref{sec:energy}
may also reduce the required efficiency.
We also calculate the multi-wavelength afterglows in
Sect.~\ref{sec:multi}.
We compare our results with observations and
discuss implications
in Sect.~\ref{sec:dis}.

\section{Efficiency crisis}\label{sec:crisis}

Let us show that a high gamma-ray efficiency is necessary to explain 
the shallow X-ray afterglows within models proposed so far.
Here we should note that the flux decay is shallower than $\propto t^{-1}$
and therefore more time-integrated energy is radiated at a later time.
In this section we assume that electrons are accelerated to a power-law
distribution $N(\gamma_e) \propto \gamma_e^{-2}$
and X-rays arise from fast cooling electrons,
so that the X-ray luminosity is proportional to 
the bolometric one.
Note that the lack of spectral evolution across the X-ray break
indicates that these breaks are not produced by the passage of a spectral
break.

(a1) First we consider the energy injection model
caused by the long-lived central engine.
If there is no injection,
the light curve decays as $\propto t^{-1}$
after the peak time
$t_{\rm dec} \sim \max[T,t_{\gamma}]$, where
$T$ is the burst duration,
\beqa
t_{\gamma} = \left(\frac{3 E_k}{256 \pi \gamma^8 n m_p c^5}\right)^{1/3}
\sim 100 E_{k,53}^{1/3} \gamma_{2}^{-8/3} n^{-1/3}\
{\rm s}
\label{eq:tg}
\eeqa
is the time to collect $\gamma^{-1}$ of the ejecta mass,
$\gamma_2=10^2 \gamma$ is the Lorentz factor of the ejecta,
$n$ is the ambient density,
and $E_k=10^{53} E_{k,53}$ erg is the afterglow energy
(Sari \cite{sari97}).
The peak time ($t_{\rm dec}\sim 1$--$10^2$ s) is typically before
the end of the shallow decay ($t_s \sim 10^3$--$10^4$ s).

If the engine continues to eject outflows 
after the prompt burst,
the outflows add energy to the external shock.
Then the afterglow decay 
becomes shallower $(\propto t^{-1/2})$ than that for no injection
$(\propto t^{-1})$.
Since the decay $\propto t^{-1/2}$ is shallower than $\propto t^{-1}$,
the time-integrated injected energy $E_{\rm inj}$ is larger than the initial 
afterglow energy $E_k$ by a factor of 
$E_{\rm inj}/E_k \sim (t_s/t_{\rm dec})^{1/2} \sim 3$--$10$.
Since the burst energy is comparable to 
the afterglow energy after injection
$E_{\gamma} \sim E_{\rm inj} \sim 3$--$10 E_k$,
the gamma-ray efficiency is corrected upward as
$\epsilon_{\gamma}=E_{\gamma}/(E_{\gamma}+E_k) \simg 75$--$90\%$.
Such a high efficiency is difficult to explain with realistic
internal shock models (Kobayashi \& Sari \cite{koba01}).

(a2) Next we consider the energy injection model
caused by a short-lived central engine 
with some ranges of Lorentz factors of ejected shells.
After the internal shocks, shells are rearranged so that
outer shells are faster and inner shells are slower.
This configuration may also occur if 
the central engine ejects faster shells earlier.
Outer shells are slowed down by making the external shock.
Once the Lorentz factor of the shocked shell drops
below that of the slower shell,
the slower shell catches up with the shocked shell,
injecting energy into the forward shock.
Thus the injection time $t_i$ of a shell with a Lorentz factor $\gamma$
is about $t_i \sim t_\gamma$
in Eq. (\ref{eq:tg}), replacing 
$E_k$ with the time-integrated injected energy.

The shallow phase continues 
until the energy in slower shells becomes less than the time-integrated 
injected energy.
Then, by equating the final time of the shallow phase $t_s \sim
10^3$--$10^4$ s with the injection time 
$t_i \sim t_{\gamma}$ in Eq. (\ref{eq:tg}),
we can estimate the Lorentz factor $\gamma_{\rm peak}$
in which most energy resides
as $\gamma_{\rm peak}\sim 30$--$50$.
Since $t_i \sim t_\gamma \propto \gamma^{-8/3}$
and the time-integrated energy grows as $\sim t^{1/2}$,
the energy distribution 
is given by $dE/d \ln \gamma \propto t^{1/2} \propto \gamma^{-4/3}$
for $\gamma > \gamma_{\rm peak}$
(Granot \& Kumar \cite{granot06a}).
Therefore, the energy in the shells with $\gamma \simg 100$ is smaller than
the total injected energy $E_{\rm inj}$ by a factor of $3$--$10$.
Now we recall that only shells with $\gamma \simg 100$ 
can make the prompt burst because of the compactness problem
(Lithwick \& Sari \cite{lithwick01}).
Then the afterglow energy remaining after the burst $E_k$
is a factor $3$--$10$ smaller than the total injected energy,
i.e., $E_{\rm inj} \sim 3$--$10 E_k$,
while the burst energy is comparable to the afterglow energy after
injection $E_{\gamma} \sim E_{\rm inj}$.
Again we find that the corrected gamma-ray efficiency
is very high,
$\epsilon_{\gamma}=E_{\gamma}/(E_{\gamma}+E_k) \simg 75$--$90\%$.

One may think that the gamma-ray efficiency is not so high
if the initial energy distribution 
$dE/d \ln \gamma$ peaks at 
$\gamma_{\rm peak}>100$ and the peak moves to
$\gamma_{\rm peak} \sim 30$ after internal shocks.
However, in order for the peak Lorentz factor $\gamma_{\rm peak}$
to move down to $\gamma_{\rm peak} \sim 30$,
the shells of $\gamma>100$ have to interact with
shells of $\gamma \sim 30$.
Since the internal shock radius is determined by the lower Lorentz
factor, the internal shocks occur deeply in the optically thick region,
and therefore we cannot avoid the compactness problem.

(b) Toma et al. (\cite{toma06}) have completely discussed the
inhomogeneous jet model.
This model also needs a high gamma-ray efficiency 
$\epsilon_{\gamma}=E_{\gamma}/(E_{\gamma}+E_k) \simg 75\%$.
Therefore, all current models face an efficiency crisis!

\section{Possible solutions}\label{sec:solutions}

\subsection{Prior activity model}\label{sec:prior}

Since the afterglow energy after the shallow phase
is more or less similar to the burst energy,
we are tempted into considering that both types of
energy have the same origin.
Then, to suppress the flux of the early afterglow,
we logically have two choices:
(A) the kinetic energy of the ejecta is not converted 
into the internal energy so much in the early phase
or (B) even if the kinetic energy is converted into
the internal energy it is not radiated away so much.
The choice (B) will be discussed in Sect.~\ref{sec:micro}.

For the choice (A) one may easily think of reducing the ambient density
$n$ because the kinetic energy is released
when the ejecta is decelerated
and the deceleration time is given by 
$t_{\rm dec} \sim t_{\gamma} \propto n^{-1/3}$
in Eq. (\ref{eq:tg}).
However in this case, the early afterglow does not show
a decaying feature but a rising one (Sari \cite{sari97}).
If we adjust the ambient density so as to have a shallow decay,
we need an unrealistic density profile that drops outward
and does not connect to the conventional density for the late afterglow.
Therefore, we cannot simply reduce the ambient density.
However, how about changing both the density and velocity of the
ambient matter?
Such a situation is not implausible if 
a prior explosion occurs before the observed prompt GRBs,
for example.
It is not unreasonable to consider such a prior activity
because the X-ray flares suggest that 
the engine activity lasts very long after the burst
(Burrows et al. \cite{burrows05}; Ioka et al. \cite{ikz05})
and it may be also present before the burst.
Actually, a sizable fraction of GRBs may have
precursor activities (Lazzati \cite{lazzati05}).
A prior activity may be also expected in the supranova model
(Vietri \& Stella \cite{vietri98}),
although the ambient matter is not relativistic in this model.
In the collapsar model such a prior activity might arise if the
main burst is produced by the fallback of the ejected matter.

\begin{figure}
  \includegraphics[width=8.5cm]{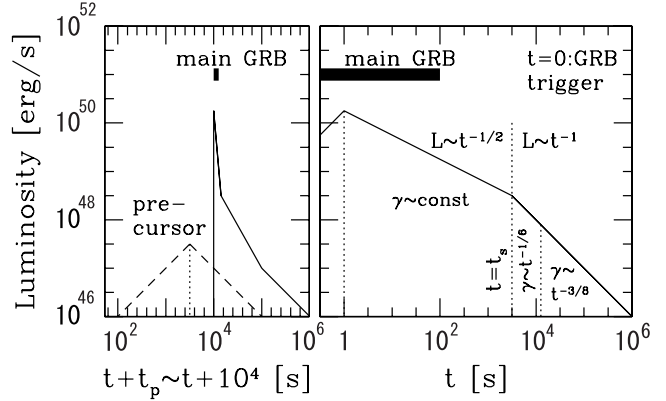}
  \caption{The afterglow luminosity as a function of time 
before the main burst ({\it left}) and 
after the main burst ({\it right})
in the prior activity model.
We set $t=0$ at the beginning of the main burst ({\it thick line}).
A precursor ({\it dashed line})
is produced by the forward shock due to the prior ejecta
launched at $t=-t_p \sim -10^4$ sec.
The forward shock emission due to the main burst ejecta
({\it solid line}) has a shallow decay $\propto t^{-1/2}$
before the burst ejecta is decelerated by the prior ejecta
$t < t_s$ in Eq. (\ref{eq:ts}),
and after that it has the conventional decay $\propto t^{-1}$.
}
  \label{fig:lc}
\end{figure}

To demonstrate the plausibility of the prior activity model,
we consider the following simple model.
We assume that an explosion occurs at $t=-t_p \sim -10^4$ s
(where we set $t=0$ as the burst trigger)
and that mass 
\beqa
M(<\gamma_p) \propto \gamma_p^{\alpha}
\label{eq:mass}
\eeqa
with Lorentz factors less than $\gamma_p $
is ejected, where
we assume $\alpha>0$ and $\gamma_p<\gamma_{\max} \sim 30$.
The energy associated with that mass is 
$E(<\gamma_p)=\gamma_p M c^2 \propto \gamma_p^{\alpha+1}$.
Since $\alpha>0$,
almost all energy is concentrated near $\gamma_{\max}$.
We also assume that a prior explosion is weaker 
than the main burst, $E(<\gamma_{\max})\equiv E_{p} 
\sim 10^{52} {\rm erg} 
<E_{\gamma} \sim 10^{53}$ erg.
The ejected mass sweeps the ambient density
making an external shock.
The deceleration begins at 
$t \sim -t_p + 10^4 E_{p,52}^{1/3} \gamma_{\max,1.5}^{-8/3} n^{-1/3}$ s
from Eq. (\ref{eq:tg}),
and then
the Lorentz factor and radius of the external shock
evolve as
\beqa
\gamma_p &\sim& 30 E_{p,52}^{1/8} n^{-1/8} [(t+t_p)/10^4\ {\rm s}]^{-3/8},
\\
R &\sim& 10^{17} E_{p,52}^{1/4} n^{-1/4} [(t+t_p)/10^4\ {\rm s}]^{1/4}
\ {\rm cm}.
\eeqa
Since the explosion is weak, its afterglow is not so bright (see
Fig.~\ref{fig:lc} and Sect.~\ref{sec:dis}).
Note that the Blandford \& McKee (\cite{bm76}) solution has
the mass profile $M(<\gamma_p) \propto \gamma_p^{3/2}$
near the shock front
and the index $\alpha$ is larger far from the shock.

We assume that the ejecta of the prompt burst at $t=0$ 
is faster than the prior ejecta, i.e.,
$\gamma>\gamma_{\max}$.
Before catching up with the external shock,
the burst ejecta will collide with the slower ejecta
at a radius $R_p \sim c t_p \gamma_p^2$.
The relative Lorentz factor between the burst and slower ejecta
is about $\eta \sim \gamma/\gamma_p$ for $\gamma \gg \gamma_p \gg 1$,
while the ratio of the comoving density is given by
\beqa
f \equiv \frac{n_k}{n_p} \sim \frac{E_k}{E(<\gamma_p)}
\propto \gamma_p^{-\alpha-1},
\eeqa
where $n_k$ ($n_p$) is the density of the burst (slower) ejecta
and where we assume the slower ejecta is cold before the collision
since it is not shocked.

Let us assume $\eta^2<f$ in the early phase.
Then the reverse shock is Newtonian (Sari \& Piran \cite{sari95}).
The burst ejecta is not decelerated and keeps its Lorentz factor $\gamma
\sim$ const.
The internal energy is mainly released in the forward shock,
which is given by
\beqa
E_i \sim M(<\gamma_p) \frac{\gamma^2}{\gamma_p}
\propto \gamma_p^{\alpha-1}.
\label{eq:ei}
\eeqa
The radiation from the collision is observed at 
\beqa
t \sim \frac{R_p}{\gamma^2} \sim t_p
\frac{\gamma_p^2}{\gamma^2}
\sim 10^3 t_{p,4}
\gamma_2^{-2}
\gamma_{p,1.5}^2
\ {\rm s},
\label{eq:t}
\eeqa
where $t_p=10^4 t_{p,4}$ s and 
$\gamma_p=10^{1.5} \gamma_{p,1.5}\sim 30 \gamma_{p,1.5}$,
and hence $\gamma_p \propto t^{1/2}$.
From Eqs. (\ref{eq:ei}) and (\ref{eq:t}),
the bolometric kinetic luminosity is given by
\beqa
L \sim \frac{E_i}{t} \propto \frac{\gamma_p^{\alpha-1}}{t} \propto
t^{(\alpha-3)/2}.
\label{eq:L}
\eeqa
Therefore, assuming that the X-ray luminosity is proportional to the
bolometric kinetic one,
we can explain the shallow decay if
$\alpha \sim 1.5$--$2.5$ (see Fig.~\ref{fig:lc}).
If the X-ray luminosity is not proportional to the
bolometric kinetic one, the light curve is given by 
Eq. (\ref{eq:X(i)})
in Sect.~\ref{sec:multi}.

For an index $\alpha > 1$,
the ratio $f/\eta^2 \propto \gamma_p^{1-\alpha}$
is a decreasing function of $\gamma_p$.
The ratio $f/\eta^2$ becomes less than unity 
as the shock expands
if
$\min (f/\eta^2) \sim \min [E_k \gamma_p^2/E(<\gamma_p) \gamma^2] \sim
E_k \gamma_{\max}^2/E_p \gamma^2 < 1$
is satisfied.
Therefore, the initially Newtonian reverse shock 
becomes relativistic.
At this point the reverse shock also crosses the burst
ejecta because the crossing radius is given 
by $R_{\Delta} \sim (f/\eta^2)^{1/2} R$ (Sari \& Piran \cite{sari95}).
Beyond the crossing radius,
we can use a simple two mass model 
to estimate the Lorentz factor of the forward shock due to the burst ejecta,
\beqa
\gamma \sim \gamma_p \left(\frac{E_k}{E(<\gamma_p)}\right)^{1/2}
\propto \gamma_p^{(1-\alpha)/2} \propto t^{(1-\alpha)/2(1+\alpha)}.
\label{eq:gf}
\eeqa
Since $\alpha>1$, the forward shock due to the burst ejecta is decelerating.
After the deceleration, the internal energy released in the forward
shock
is comparable to the energy of the burst ejecta $\sim E_{k}$.
Then the bolometric kinetic luminosity evolves as
\beqa
L \sim E_{k}/t \propto t^{-1},
\label{eq:L-1}
\eeqa
which is the conventional decay after the shallow phase.
The time when the conventional decay begins
is estimated from Eq. (\ref{eq:t}) 
with $\gamma_p$ satisfying $f/\eta^2 \sim 1$, i.e.,
\beqa
t_s \sim t_p \frac{E(<\gamma_p)}{E_k}
\siml 10^3 t_{p,4} E_{p,52} E_{k,53}^{-1}
\ {\rm s}.
\label{eq:ts}
\eeqa
This is about $\sim 10^3$ s for our parameters 
and reproduces the observations (see Fig.~\ref{fig:lc}).

Finally, the forward shock due to the burst ejecta
overtakes the outermost forward shock due to the prior ejecta.
At this radius the Lorentz factor $\gamma$
of the forward shock due to the burst
ejecta in Eq. (\ref{eq:gf})
is the same as that in the absence of the prior explosion.
Therefore we have the same luminosity evolution 
in Eq. (\ref{eq:L-1}) after the forward shock
drives into the ambient medium.
(For more strict arguments, calculations like those 
in Zhang \& {\meszaros} (\cite{zhang02}) are necessary.)
Since the prior explosion has less energy than the prompt burst,
the final afterglow energy is comparable to the initial afterglow energy
$E_{k}$, and hence we have no efficiency crisis.

In summary,
a shallow light curve can be reproduced without the efficiency crisis
if a small explosion with less energy than the main burst
occurs $t_p \sim 10^3$--$10^6$ s before the burst.
The shallow phase ends at around $t_s \sim 10^3$--$10^4$ s 
in Eq. (\ref{eq:ts}),
and this time marks the beginning of the
deceleration of the burst ejecta due to the prior ejecta.
The decay index of the shallow phase in Eq. (\ref{eq:L})
is mainly determined by
the mass distribution of the prior ejecta in Eq. (\ref{eq:mass}).
More detail calculations are given in Sect.~\ref{sec:multi}
and are compared with observations in Sect.~\ref{sec:dis}.

\subsection{Time-dependent microphysics model}\label{sec:micro}

The other possibility to obtain the shallow X-ray afterglow 
without the efficiency crisis is
to vary the microphysical parameters,
such as the energy fraction that goes into electrons $\epsilon_e$
and magnetic fields $\epsilon_B$, during the observations.
Even if the burst ejecta is decelerated and 
the internal energy is released,
most internal energy is initially carried by protons.
Without transferring the proton energy into electrons and magnetic
fields,
little radiation is emitted since
protons are inefficient emitters.

So far we usually assume that the microphysical parameters are not
varying and in fact, constant $\epsilon_e$ and $\epsilon_B$
are consistent with the observations of late afterglows
(Yost et al. \cite{yost03}).
However, since the first few hours after the burst
is an unexplored period,
we should check the constancy of microphysics observationally 
in this time interval without having any prejudice.
Also, on the theoretical side,
the mechanism of the energy transfer from protons to electrons
and magnetic fields in the relativistic shocks
is not well understood from the first principles.
Although recent particle simulations have demonstrated that the
magnetic fields are generated by the Weibel instability
in collisionless shocks
(Medvedev \& Loeb \cite{medvedev99}; Silva et al. \cite{silva03};
Kato \cite{kato05}),
the long term evolution up to the time of the actual observation
is beyond the current computer power.
In addition, simulations have not succeeded
in reproducing $\epsilon_e \sim 0.1$ 
(but see Hededal \cite{hededal05}),
probably because 
the grid size is not small enough for resolving 
the radiation wavelength, and hence the coherent effects
are not properly calculated
(Ioka \cite{ioka05}).
Since the coherent effects could depend on the 
Lorentz factor of the shock (Ioka \cite{ioka05}),
the electron energy fraction may vary
in the early afterglow.

If the index of the power-law electron distribution $p$
is about $p \sim 2$ as usual
and fast cooling electrons emit X-rays,
the X-ray luminosity $L_X$ is given by the bolometric kinetic luminosity
$L$ as
\beqa
L_{X} \sim \epsilon_e L,
\eeqa
and does not depend on the magnetic energy fraction $\epsilon_B$
so much ($L_X \propto \epsilon_B^{(p-2)/4}$).
Since $L \propto t^{-1}$, 
the shallow X-ray light curve $L_X \propto t^{-1/2}$ suggests that
the electron energy fraction evolves as
\beqa
\epsilon_e \propto t^{1/2},
\eeqa
which is saturated at the equipartition value $\epsilon_e \sim 0.1$--$1$
when the shallow phase ends.
Note that the initial value of $\epsilon_e$ at $t \sim 1$--$100$ s
is still larger than the minimum energy fraction
$\epsilon_{e,\min}=m_e/m_p \sim 10^{-3}$.
More detail calculations are presented in Sect.~\ref{sec:multi}.

\subsection{Energetic afterglow model}\label{sec:energy}

The gamma-ray efficiency $\epsilon_{\gamma}=E_\gamma/(E_\gamma+E_k)$
can be reduced if
the kinetic energy of the afterglows $E_k$ is actually much larger 
than previously thought (Fan \& Piran \cite{fan06};
Granot, Konigl, \& Piran \cite{granot06b}).
This could be caused by the following possibilities.
(i) The standard afterglow model has some ambiguities.
In particular, Lloyd-Ronning \& Zhang (\cite{lloyd04}) take a relatively
large typical synchrotron frequency $\nu_m$, which increases
$\epsilon_{\gamma}$.
(ii) Almost all energy could be radiated by
the inverse Compton emission that is not observed.
(iii) Only a part of the electrons $\zeta_e$ could be accelerated
(Papathanassiou \& {\meszaros} \cite{papa96}).
Thus, the afterglow fitting could actually give a larger
kinetic energy (Eichler \& Waxman \cite{eichler05}).

Regardless, in this model the kinetic energy of the afterglow should be 
$\sim 10$ times larger than previously thought 
for a moderate gamma-ray efficiency $\epsilon_\gamma \sim 0.1$.
In other words, the total energy of the gamma-ray bursts
is $\simg 10^{52}$ erg rather than $\sim 10^{51}$ erg.
Note that this model itself does not explain the shallow decay 
of the X-ray afterglow. 
We also have to consider time-dependent microphysical parameters
($\epsilon_e, \epsilon_B, \zeta_e$), the energy injection,
or the inhomogeneous jet.

\section{Multi-wavelength afterglows}\label{sec:multi}

First, let us consider the prior activity model.
There are three phases in this model.
(i) The first one is the $\gamma \sim$ const. phase, which corresponds to
the shallow X-ray phase.
(ii) The second one follows after the reverse shock crosses the burst ejecta.
(iii) Finally, the conventional evolution in the constant density medium begins
after the burst ejecta overtakes the outermost
forward shock due to the prior ejecta.
In the first phase (i),
we estimate the comoving density of the pre-shocked matter $n_p$,
the typical Lorentz factor of shocked electrons $\gamma_m$,
the comoving magnetic field in the shocked region $B$,
and the total number of emitting electrons $N_e$ as
\beqa
n_p &=& \frac{M(<\gamma_p)}{4\pi m_p R_p^3/\gamma_p}
\propto \gamma_p^{\alpha-5} \propto t^{(\alpha-5)/2},
\label{eq:np}
\\
B &=& (32\pi m_p \epsilon_B n_p)^{1/2} \eta c
\propto \epsilon_B^{1/2} t^{(\alpha-7)/4},
\\
\gamma_m &=& \epsilon_e \frac{p-2}{p-1} \frac{m_p}{m_e} \eta
\propto \epsilon_e \gamma_p^{-1} \propto \epsilon_e t^{-1/2},
\\
\gamma_c &=& \frac{6\pi m_e c}{\sigma_T \gamma B^2 t}
\propto \epsilon_B^{-1} t^{(5-\alpha)/2},
\\
N_e &=& \frac{M(<\gamma_p)}{m_p}
\propto \gamma_p^{\alpha} \propto t^{\alpha/2}.
\eeqa
Then, according to the standard afterglow theory 
(Sari, Piran \& Narayan \cite{sari98}),
we can calculate the spectral evolutions as
\beqa
\nu_m &=& \frac{\gamma \gamma_m^2 q B}{2\pi m_e c} \sim
1 \times 10^{12}{\rm Hz}\ \epsilon_{e,-1}^{2} 
\epsilon_{B,-2}^{1/2} t_3^{(\alpha-11)/4},
\\
\nu_c &=& \frac{\gamma \gamma_c^2 q B}{2\pi m_e c} \sim
8 \times 10^{15}{\rm Hz}\ \epsilon_{B,-2}^{-3/2}
t_3^{(13-3\alpha)/4},
\\
F_{\nu,{\rm max}} &=& \frac{N_e m_e c^2 \sigma_T \gamma B}
{12 \pi D^2 q} \sim 
0.4 {\rm Jy}\ \epsilon_{B,-2}^{1/2} t_3^{(3\alpha-7)/4},
\label{eq:fnumax}
\eeqa
where $\epsilon_{e,-1}=\epsilon_e/0.1$,
$\epsilon_{B,-2}=\epsilon_B/0.01$,
$t_3=t/10^3$ s,
and we adopt $p=2.2$,
$\gamma=100$,
$\gamma_{\max}=30$,
$E_p=10^{52}$ erg,
$t_p=10^4$ s, and $\alpha=2$.
Therefore, the X-ray light curve is given by
\beqa
F_{\nu_c<\nu}&=&F_{\nu,\max} (\nu_c/\nu_m)^{(1-p)/2}
(\nu/\nu_c)^{-p/2}
\nonumber\\
&\propto& t^{\frac{\alpha-3}{2}+\frac{(\alpha-11)(p-2)}{8}},
\label{eq:X(i)}
\eeqa
which is consistent with Eq. (\ref{eq:L})
since $F_{\nu_c<\nu} \propto t^{(\alpha-3)/2}$ for $p=2$.
(Note that the luminosity per logarithmic frequency 
is constant for $p=2$ and so
the X-ray luminosity is proportional to the bolometric one.)
The optical light curve is
\beqa
F_{\nu_m<\nu<\nu_c}=F_{\nu,\max} (\nu/\nu_m)^{(1-p)/2}
\propto t^{\frac{7\alpha-25}{8}+\frac{(\alpha-11)(p-2)}{8}}.
\label{eq:O(i)}
\eeqa

After the reverse shock crosses the burst ejecta
in the prior activity model (i.e., in phase (ii)),
the Lorentz factor of the forward shock evolves 
according to Eq. (\ref{eq:gf}),
and so we have $\gamma_p \propto t^{1/(1+\alpha)}$
with Eq. (\ref{eq:t}).
By using the same Eqs. (\ref{eq:np})-(\ref{eq:fnumax})
with different $t$ dependences of $\gamma$ and $\gamma_p$,
we can find
\beqa
\nu_m \propto t^{(-3\alpha-7)/2(1+\alpha)},
\\
\nu_c \propto t^{(-3\alpha+13)/2(1+\alpha)},
\\
F_{\nu,\max} \propto t^{(\alpha-5)/2(1+\alpha)}.
\eeqa
Therefore, the X-ray light curve is
\beqa
F_{\nu_c<\nu} \propto t^{-1-\frac{(3\alpha+7)(p-2)}{4(1+\alpha)}},
\label{eq:Xii}
\eeqa
which is consistent with Eq. (\ref{eq:L-1}), since 
$F_{\nu_c<\nu} \propto t^{-1}$ for $p=2$.
The optical light curve is
\beqa
F_{\nu_m<\nu<\nu_c} \propto 
t^{-\frac{\alpha+17}{4(1+\alpha)}-\frac{(3\alpha+7)(p-2)}{4(1+\alpha)}}.
\label{eq:Oii}
\eeqa
After the burst ejecta runs into the interstellar medium 
(i.e., in phase (iii)),
we note the conventional light curve as
\beqa
F_{\nu_c<\nu} \propto t^{-1-3(p-2)/4},
\quad
F_{\nu_m<\nu<\nu_c} \propto t^{-3(p-1)/4}.
\label{eq:SPN}
\eeqa

The reverse shock emission 
in the prior activity model
peaks when the shock crosses the burst ejecta
in Eq. (\ref{eq:ts}).
Since the reverse shock is relativistic at $t\sim t_s$,
the energy and energy density are similar to that of the forward shock.
The thermal Lorentz factor is $\eta^{-1}$ times smaller,
while the electron number is $\eta$ times larger
than that of the forward shock.
Then $\nu_m$, $\nu_c$, and $F_{\nu,\max}$
are $\eta^{-2}$, $1$, and $\eta$ times larger than
that of the forward shock, respectively.
Therefore, the reverse shock emission is dominated by the forward shock
emission, at least in the optical band.

Second, let us consider the time-dependent microphysics model.
For general arguments, we parameterize
\beqa
\epsilon_e \propto t^{\beta_e}, \quad \epsilon_B \propto t^{\beta_B},
\label{eq:eeeb}
\eeqa
in the shallow X-ray phase.
Since the X-ray band is typically above $\nu_c$ and
the optical band is between $\nu_m$ and $\nu_c$,
the X-ray and optical light curves are given by
\beqa
F_{\nu_c<\nu} &\propto&
t^{-1+\beta_e+\frac{(4\beta_e+\beta_B-3)(p-2)}{4}},
\label{eq:Xt}
\\
F_{\nu_m<\nu<\nu_c} &\propto& t^{\frac{4\beta_e+3\beta_B-3}{4}
+\frac{(4\beta_e+\beta_B-3)(p-2)}{4}},
\label{eq:Ot}
\eeqa
respectively.
After the shallow X-ray phase, the optical light curve returns to
the conventional decay in Eq. (\ref{eq:SPN}).
The reverse shock emission is dominated by the forward shock
emission in the optical band, since $\nu_m$, $\nu_c$, and $F_{\nu,\max}$
are $\gamma^{-2}$, $1$, and $\gamma$ times larger than
that of the forward shock, respectively,
if the microphysical parameters evolve in a similar manner to
the forward shock.
We do not consider the energetic afterglow model in this Letter
since the model itself does not explain the shallow X-ray decay.

\section{Discussion}\label{sec:dis}

We have discussed that the shallow decay of the X-ray afterglow
requires an unreasonably high gamma-ray efficiency, $\simg 75-90\%$,
within current models (the energy injection model
and the inhomogeneous jet model).
Since such an efficiency is difficultly produced by internal shocks,
we have suggested other possibilities, i.e., the prior activity model
and the time-dependent microphysics model,
that have no efficiency crises.
We have also calculated the multi-wavelength afterglows 
in these models.

Let us compare the multi-wavelength afterglows with observations,
although the multi-wavelength observations of the early afterglows
are not yet very abundant.
For several events,
the X-ray and optical decay indices are
around $-(0.5$--$1)$ and $-(0.4$--$0.8)$, respectively,
in the shallow X-ray phase (Panaitescu et al. \cite{pana06b}).
In the prior activity model,
such behavior occurs if
$p\sim 3$ (the power-law index of the electron distribution)
and $\alpha \sim 3.8$ (in Eq. (\ref{eq:mass})), which give
indices of $\sim -0.5$ (X-ray) and $\sim -0.7$ (optical)
from Eqs. (\ref{eq:X(i)}) and (\ref{eq:O(i)}).
However, the spectral index $-p/2 \sim -1.5$
in Eq. (\ref{eq:X(i)}) is inconsistent 
with the observations ($\sim 1$).
Thus, this model should be modified,
e.g., by making microphysical parameters time-dependent.
In the time-dependent microphysics model,
we have the observed indices
$\sim -0.5$ (X-ray) and $\sim -0.7$ (optical)
for $p\sim 2$, $\beta_e \sim 0.5$, and $\beta_B \sim -0.6$
from Eqs. (\ref{eq:eeeb}), (\ref{eq:Xt}), and (\ref{eq:Ot}).
(We adopt $p\sim 2$ to fit the post-shallow phase.
See below.)
Thus, the magnetic field energy also decreases during the shallow phase.
It is interesting to note that the particle simulations
also suggest a long term decay of magnetic fields.
(Silva et al. \cite{silva03}; Kato \cite{kato05}).

The multi-wavelength observations of the early afterglows also
suggest that the shallow X-ray phase ends without any optical break
(Panaitescu et al. \cite{pana06b}).
In the prior activity model,
the X-ray and optical decay indices
are $\sim -2$ and $\sim -2$, respectively, 
just after the shallow X-ray phase
from Eqs. (\ref{eq:Xii}) and (\ref{eq:Oii}),
and finally $\sim -1.7$ and $\sim -1.5$, respectively,
from Eq. (\ref{eq:SPN}),
if we use the same parameters, $p\sim 3$ and $\alpha \sim 3.8$,
as in the shallow X-ray phase.
This is inconsistent with the observations.
To explain the observations in this model,
the power-law index of the electron distribution $p$
should finally change to the conventional value $p\sim 2$.
In the time-dependent microphysics model,
the chromatic X-ray break can be explained 
if the microphysical parameters stop evolutions
after the shallow X-ray phase
since the X-ray and optical decay indices
become $\sim -1$ and $\sim -0.8$, respectively, for $p\sim 2$
from Eq. (\ref{eq:SPN}).
Note that the simplest form of the current models
(the energy injection and the inhomogeneous jet model)
predicts an optical break at the end of the shallow X-ray decay
because the hydrodynamics changes.

The prediction of the prior activity model is
a precursor produced by the external shock due to the prior explosion
(see Fig.~\ref{fig:lc}).
Such a precursor may have evaded the detection 
since its luminosity could be low if the maximum Lorentz factor
of the prior explosion $\gamma_{\max}$ is not so large.
The prompt emission from the prior explosion may be also dim
if $\gamma_{\max}$ is too low to avoid the compactness problem.
The precursor emission peaks around the deceleration time
$t_{\rm dec} \sim 10^4 E_{p,52}^{1/3} 
\gamma_{{\rm max},1.5}^{-8/3} n^{-1/3}$ s in Eq. (\ref{eq:tg}),
and it is $t_p-t_{\rm dec} \sim 10^3$--$10^6$ s
before the main burst.
The peak luminosity of the precursor is about
$L_p \sim \epsilon_e E_p/t_{\rm dec} \sim 10^{47} \epsilon_{e,-1}
E_{p,52} t_{{\rm dec},4}^{-1}$ 
erg s$^{-1}$.

In both the prior activity model
and the time-dependent microphysics model, 
it is predicted that
the optical emission from the reverse shock is suppressed
(see Sect.~\ref{sec:multi}).
%This is because it takes longer time for 
%the reverse shock to become relativistic
%than usually considered.
This may be relevant to the dim optical flashes
from the reverse shock recently reported (Roming et
al. \cite{roming05}).
A possibility that the prior activity is continuous
is also interesting to study.

%The predictions of the time-dependent microphysics model 
%is that in the beginning of the afterglow
%the typical synchrotron frequency 
%$\nu_m \propto \epsilon_e^{2} t^{-3/2}$ 
%is relatively low (possibly below optical)
%and it evolve as $\nu_m \propto t^{-1/2}$ in a similar fashion to the
%cooling frequency $\nu_c \propto t^{-1/2}$ 
%(Sari, Piran \& Narayan \cite{sari98}).
%A multi-wavelength observations are useful to test this model.
%It is also interesting to study the time-dependent microphysics model
%in the reverse shock, which may explain the dim optical flashes
%from the reverse shock (Roming et al. \cite{roming05}).

\begin{acknowledgements}
We thank Y.~Z.~Fan and C.~Hededal for useful comments.
This work is supported in part 
by Grant-in-Aid for the 21st Century COE
``Center for Diversity and Universality in Physics''
from the Ministry of Education, Culture, Sports, Science and Technology
(MEXT) of Japan and also 
by Grants-in-Aid for Scientific Research
of the Japanese Ministry of Education, Culture, Sports, Science,
and Technology 18740147 (K.I.), 14047212 (K.~I. and T.~N.), 
14204024 (K.~I. and T.~N.), 18740153 (R.Y.),
and 17340075 (T.~N.).
\end{acknowledgements}

%\appendix

\end{document}